\documentclass[final,3p,times]{elsarticle}

\usepackage{epsfig}

\usepackage{amssymb}
\usepackage{amsmath}
 \usepackage{amsthm}
\usepackage{graphicx}
\usepackage[colorlinks=true, urlcolor=blue, linkcolor=red]{hyperref}
\usepackage[numbers]{natbib}

\journal{Chinese Journal of Physics}

\begin{document}

\begin{frontmatter}



\title{Determination of Asymptotic Normalization Coefficients Based on the Dispersive Optical Model}

\author[1]{O. V. Bespalova}
\author[1]{L.D. Blokhintsev}
\author[1,2]{A.A.Klimochkina\corref{cor1}}
\ead{klimann16@gmail.com}
\author[1]{D.A. Savin}

\cortext[cor1]{Corresponding author}

\affiliation[1]{organization={Skobeltsyn Institute of Nuclear Physics, Lomonosov Moscow State University},
            city={Moscow},
            postcode={119991}, 
            country={Russia}}
\affiliation[2]{organization={Faculty of Physics, Lomonosov Moscow State University},
             city={Moscow},
            postcode={119991},
             country={Russia}}
\begin{abstract}
	A method for determining asymptotic normalization coefficients for removing nucleons from nuclei is proposed. It is based on the use of the dispersive optical model potential. Within the method, the strength parameter of the Hartree-Fock type potential at the Fermi energy is the only one fitting parameter. It was found that adjusting this parameter allows us to achieve a coincidence of the calculated binding energies  with the experimental ones accurately enough. Specific calculations were carried out for $^{17}$O, $^{17}$F, $^{41}$Ca, and $^{41}$Sc nuclei. An acceptable agreement with the results of other works, which are characterized by a noticeable spread,  was achieved.
\end{abstract}


\end{frontmatter}

\section{Introduction}

	Asymptotic normalization coefficients (ANCs) determine the asymptotics of bound-state nuclear wave functions in binary channels.  
 They are fundamental nuclear characteristics which are important both in nuclear reaction and nuclear structure physics. ANCs are used actively in analyses of nuclear reactions within various approaches, in particular, in the distorted wave Born approximation (DWBA). The ANCs extracted from the analysis of one process can be used to predict features of other processes. Comparing empirical values of ANCs with theoretical ones enables one to evaluate the  possibilities of a model. 

The ANC for the $a\to b+c$ channel determines the probability of the $b+c$ configuration in nucleus $a$ at distances greater than the radius of the nuclear interaction.  That is why  
ANCs are especially important for the analysis of peripheral nuclear processes, when the reaction occurs at large distances between fragments due to the presence of a large Coulomb (or centrifugal) barrier. It is natural to parameterize such processes in terms of the ANCs, rather than the commonly used spectroscopic factors. The most important representative of such processes are astrophysical nuclear reactions occurring in the cores of stars. A significant part of these reactions is inaccessible for direct measurement in laboratory conditions due to the smallness of their cross sections caused by the large Coulomb barrier. This circumstance increases the importance of knowledge of the ANCs. 

The role of ANCs in nuclear astrophysics was first discussed in \cite{Mukh1,Mukh6}
 where it was emphasized that they determine the overall normalization of peripheral radiative capture reactions.  ANCs are on-shell quantities similar to reaction cross sections and phase shifts. Therefore, it is natural that information about ANCs can be obtained from experimental data on nuclear reactions. However, unlike binding energies, ANCs cannot be measured directly, and special methods are required to extract them from experimental data. 
 In particular, the values of ANCs can be extracted by comparing the absolute values of experimental cross sections of nuclear transfer reactions with theoretical ones calculated within the DWBA.
Some other methods for determining ANCs as well as theoretical calculations of ANCs are discussed in the review \cite{MukhBlokh}. 

In the present paper, the ANCs were calculated within the dispersive optical model (DOM)  \cite{Mah91}. In the model, the coupling of single-particle motion with more complex configurations is taken into account by the dispersive component of the dispersive optical model potential (DOMP).  The DOM mean field is unified for negative and positive energies so that the properties of the nucleon bound states and nucleon scattering cross sections can be described by the unified manner. Very good agreement of the calculated nucleon scattering cross sections and single-particle properties of the stable nuclei with the experimental data was achieved for nuclei at mass number $40 \leq A \leq 208$ and energy  $ 70 \leq E \leq 200$ MeV using DOMP \cite{Mah91,Muller2011,BR15,Dick17,Cap05}. The DOM provides us with the possibility to trace the evolution of the nuclear single-particle structure of nuclei when $N$ and $Z$ numbers change over a wide region \cite{Muller2011,BR15}.  DOM was applied to (d,p) transfer reactions on closed-shell nuclei and ANCs  for separation of a neutron from  $^{41,49}$Ca, $^{133}$Sn, and $^{209}$Pb nuclei were determined  \cite{Nguyen}. 
The paper is organized as follows. Section 2 describes the general formalism of the method used. Section 3 is devoted to the application of this method to the determination of the ANCs for removing nucleons from $^{17}$O, $^{17}$F, $^{41}$Ca, and $^{41}$Sc nuclei. The obtained results are briefly discussed and compared with the results of other works in Section 4.

\section{Basic formalism}

	\subsection {Asymptotic normalization coefficients} 

	In this subsection we give the definition of ANCs. Throughout the paper  we use the system of units in which $\hbar=c=1$.  More detailed information about ANCs can be found in review papers \cite{MukhBlokh,BBD}.

	Let us consider  the composite bound system $a$, which can be divided into two subsystems (fragments) $b$ and $c$. The overlap function 
(aka the overlap integral) $I(\mathbf{r})$ is defined as 
\begin{equation}\label{overlap}
I_{abc}(\mathbf{r})=\int\psi_b^+(\tau_b)\psi_c^+(\tau_c)\psi_a(\tau_b,\tau_c,\mathbf{r})d\tau_bd\tau_c,
\end{equation}
where $\psi_i(\tau_i)$ is the inner wave function of system $i$ depending on inner coordinates $\tau_i$ and $\mathbf {r}$ is the radius-vector connecting the centres-of-mass of $b$ and $c$.
 In fact, $I_{abc}(\mathbf {r})$ is the projection of the wave function of $a$ onto the $b+c$ channel. It is common knowledge that overlap functions appear as important parts of matrix elements within various approaches to describing nuclear reactions. The overlap function (\ref{overlap}) can be represented as a partial-wave decomposition:
\begin{equation}\label{overlap1}
I_{abc}(\mathbf{r})=\sum_{lsm_lm_s}i^l\,(J_bM_bJ_cM_c|sm_s)\,(lm_lsm_s|J_aM_a)Y_{lm_l}(\mathbf{r}/r)I_{abc}(ls;r),
\end{equation}
where $l$ ($s$) is the channel orbital momentum (channel spin) and $I_{abc}(ls;r)$ is the radial overlap function. 

Note that $I_{abc}(ls;r)$ is normalized not to unity but to the spectroscopic factor $S_{abc}(ls)$:
\begin{equation}\label{spectr}
\int_0^\infty |I_{abc}(ls;r)|^2r^2dr=S_{abc}(ls).
\end{equation}
It is generally supposed that for short-range (nuclear) forces 
\begin{equation}\label{asympt}
I_{abc}(ls;r)|_{r>R_N}=C_{abc}(ls)\sqrt{\frac{2\kappa}{\pi r}} K_{l+1/2}(\kappa r)\longrightarrow C_{abc}(ls)e^{-\kappa r}/r\quad \mathrm{at}\quad r\to\infty.
\end{equation}
In Eq. (\ref{asympt}), $K_{l+1/2}(z)$ is the modified Hankel function, $\kappa^2=2\mu_{bc}\varepsilon$, $\mu_{bc}$ is the reduced mass of $b$ and $c$ fragments, $\varepsilon=m_b+m_c-m_a$ is the binding energy of $a$ in the $b+c$ channel, $m_i$ is the mass of particle $i$, and $R_N$ is the radius of nuclear interaction.

$C_{abc}(ls)$ is called the asymptotic normalization coefficient, it has the dimension fm$^{-1/2}$. Along with $C_{abc}(ls)$, sometimes dimensionless ANCs $\bar C_{abc}(ls)$  are also used, related to $C_{abc}(ls)$ as $C_{abc} = \sqrt{2\kappa}\bar C_{abc}$. In the case of a zero-range potential, $\bar C_{abc}=1$. In the following, for brevity we will use the notation 
$C_l$ for the ANC $C_{abc}(ls)$, $S_l$ for the spectroscopic factor $S_{abc}(ls)$ and $I_l(r)$ for the radial overlap function 
$I_{abc}(ls;r)$.  

It should be noted that the asymptotic form (\ref{asympt}) is strictly true only for two-body systems for which it follows directly from the two-body Schr\"odinger equation. For a many-body system $a$, in rare cases the asymptotics of 
$I_{abc}(ls;r)$ may differ from that of Eq. (\ref{asympt}) (the so-called anomalous asymptotics \cite{Blokh1,Blokh2}).  ANC $C_l$ is proportional to the nuclear vertex constant $G_l$ which coincides with the partial-wave matrix element of the virtual process {$a\to b+c$ on-shell \cite{BBD}.

A partial-wave amplitude  $f_l$ of the $b+c$ elastic scattering has a pole in the c.m.s. energy $E$ at  $E=-\varepsilon$ corresponding to the bound state $a$.  The residue of $f_l$ at this pole is expressed in terms of the ANC $C_l$: 
\begin{equation}\label{res}
\mathrm {res}f_l(E)|_{E=-\varepsilon}=\lim_{\substack{E\to -\varepsilon}}[(E+\varepsilon)f_l(E)] =-\frac{1}{2\mu}C_l^2.
\end{equation}
Eq. (\ref{res}) is a very important relation since it connects the characteristics of bound and continuum states. 

If both particles $b$ and $c$ are charged, the asymptotics (\ref{asympt}) of $I_{abc}(ls;r)$ is modified
\begin{equation}\label{asympt1}
I_l(r)|_{r>R_N}=C_l\frac{W_{-\eta,l+1/2}(2\kappa r)}{r}\rightarrow C_l\frac{\exp(-\kappa r)}{(2\kappa r)^{\eta}r}\quad \mathrm{at}\quad r\to\infty. 
\end{equation}
In \eqref{asympt1}, $W_{\alpha,\beta}(z)$ is the Whittaker function, $\eta=Z_bZ_ce^2\mu_{bc}/\kappa$ is the Coulomb (Sommerfeld) parameter for the bound state $a$ in the $b+c$ channel, $Z_ie$ is the electric charge of particle $i$. This equation follows directly from the 
Schr\"odinger equation containing the Coulomb potential $V_{bc}=Z_bZ_ce^2/r$.
In the presence of the Coulomb interaction Eq. (\ref{res}) holds for the Coulomb--nuclear amplitude. 

Overlap functions naturally arise in expressions for matrix elements of nuclear transfer reactions. However, overlap functions are solutions of the many-body problem. Therefore, when describing data on transfer reactions within the framework of the DWBA, overlap functions are traditionally replaced by single-particle bound-state wave functions, which are solutions of the single-particle Schr\"odinger equation with optical potential.	

\subsection {Dispersive optical model potential } 

 The DOM \cite{Mah91} describes the scattering of a nucleon by a target nucleus with $A$ nucleons and quasiparticle excitations in both the $(A + 1)$- and $(A - 1)$-nucleon systems by a unified potential.  To calculate a wave function and an energy of a nucleon bound state, the Schr\"{o}dinger equation is solved with the real part of the dispersive optical model potential. The  central real part of the DOMP for neutrons can be expressed as the sum of the  Hartree-Fock  type component  $V$\raisebox{-.4ex}{\scriptsize HF} and dispersive components $\Delta V$:
\begin{eqnarray} \label{intro}
V(r,E)=-V_{\textrm {HF}}(E)f(r,r_{\textrm {HF}},a_{\textrm {HF}})- \nonumber \\
-\Delta V_{s}(E)f(r,r_{s},a_{s})+4a_{d}\Delta V_{d}(E)\frac{d}{dr}f(r,r_{d},a_{d}),
  \end{eqnarray}
where $f(r,r_i, a_i)$ is the Woods-Saxon function, indexes $i = $HF, $ s, d$ refer to the Hartree-Fock  type component, volume and surface dispersive components, respectively. For protons, the Coulomb potential was added in the form of a uniformly charged sphere of the radius $R=A^{1/3}r_{\textrm{C}}$. 
 
The dispersive components  $\Delta V_{s,d}$ arise from the dispersion relation which connects the real and imaginary parts of the DOMP: 

\begin{eqnarray}
\Delta V_{s(d)}(E)=\frac{P}{\pi}\int\limits_{-\infty}^{\infty}\frac{W_{s(d)}(E')}{(E'-E)}dE' ,
\end{eqnarray}
where $P$ stands for the principal value. Dispersive components were calculated analytically according to \cite{Vandercam}. For this, the strength parameter $W_{s,d}(E)$ of the imaginary part was assumed to be symmetric concerning the Fermi energy and was approximated at $E \ge E_\textrm F$ by the expressions:
\begin{displaymath}
W_s(E) = \left\{ \begin{array}{ll}
0 & \textrm{ $E_F \leq E < E_p$}\\
w_1 \frac{(E-E_p)^2}{(E-E_p)^2+(w_2)^2} & \textrm{$E \geq E_p$}                        ,\\
\end{array} \right.
\end{displaymath}

\begin{displaymath}
W_d(E) = \left\{ \begin{array}{ll}
0 & \textrm{ $E_F \leq E < E_p$}\\
d_1 \frac{(E-E_p)^2 exp (-d_2(E-E_p))}{(E-E_p)^2+(d_3)^2} & \textrm{$E \geq E_p$}                     .\\
\end{array} \right.
\end{displaymath}
The energy $E_{ \textrm F}$ was found using empirical data on the nucleon separation energies $S_{n(p)}$ from nuclei with mass numbers $A$ and    $A+1$ \cite{AME}:
\begin{eqnarray}\label{Ef}
E_{F}=-\frac{1}{2}(S_{n(p)}(A)+S_{n(p)}(A+1)).
\end{eqnarray}
The energy $E_{p}$ (\ref{Ef}) is related to the experimental particle-hole energy gaps. It was evaluated following \cite{Muller2011}.  The KD global parameters were used when choosing $w _{1} , w _{2} , d_{1} , d_{2} , d_{3}$. In the present paper, the parameters of the spin-orbital and Coulomb parts of the DOMP were taken from the traditional (non-dispersive) KD optical model potential \cite{KD} as well as the geometrical parameters  $r_{ \textrm {HF}}=r_{s}$, $a_{ \textrm{HF}}=a_{s}$, $r_{d}$, $a_{d}$. The parameter $a_{ \textrm{HF}}$ = 0.58 fm for n, p -$^{16}$O systems was an exception, it was chosen to describe the particle-hole gaps in $^{16}$O. The use of global parameters KD provides an average description of the nucleon scattering data for nuclei under study.

Within the DOM, the radial wave function $u_{nlj}(r)$ which is the solution of the Schr\"{o}dinger equation, is corrected to take into account the nonlocality effect:
\begin{eqnarray}
\bar{u}_{nlj}=c_{nlj}(m^{*}_{ \textrm{HF}}(r,E)/m)^{1/2}u_{nlj}(r) .
\end{eqnarray}
The ratio  of the Hartree-Fock effective mass of a nucleon $m_{ \textrm{HF}}^{*}(r,E)$ to its free mass $m$ is given by the expression:
\begin{eqnarray}
m^{*}_{\textrm{HF}}(r,E)/m=1-\frac{d}{dE}V_{ \textrm{HF}}(r,E) .
\end{eqnarray}
The coefficient $c_{nlj}$ is determined by normalizing $\bar{u}_{nlj}(r)$ to unity:
 \begin{eqnarray} \label{baru}
\int_0^\infty |\bar{u}_{nlj}(r)|^2dr=1.
\end{eqnarray}
The DOM, unlike the traditional optical model, allows one to find not only wave functions, but also the corresponding spectroscopic factors $S_{nlj}$ using the formula:
\begin{eqnarray}\label{spectr1}
S_{nlj}=\int_0^{\infty}\bar{u}_{nlj}^2(r)\Bigl[\frac{m}{\bar{m}(r,E_{nlj})}\Bigr]dr,
\end{eqnarray}
where $\bar{m}(r,E)$ is the energy-dependent effective mass which in turn is related to the dispersive component of DOMP by the expression :
\begin{eqnarray}
\frac{\bar{m}(r,E_{nlj})}{m} = 1 - \Bigl[\frac{m}{m^*_{\textrm{HF}}(r, E)} \Bigr]\frac{d}{dE}\Delta V(r,E).
\end{eqnarray}

The energy dependence of the strength parameter $V$\raisebox{-.4ex}{\scriptsize HF} of the HF component (\ref{intro}) is smooth and arises due to the transition from a non-local potential to a local one. In the present paper, it is parametrized as following:
\begin{eqnarray}
V_{\textrm{HF}}(E)=V_{\textrm{HF}}(E_\textrm F) \exp\biggl(\frac{-\gamma(E-E_{\textrm F})}{V_{\textrm {HF}}(E_{\textrm F})}\biggr).
\end{eqnarray}
The accuracy of $\gamma$ is not so important in the present study because this parameter affects the energy $E_{nlj}$ of the deep bound states to a greater extent than on the energy $E_{nlj}$ of the states near the Fermi energy. The value $\gamma =  0.44 $ was chosen for all of the nuclei. So, the parameter  $V$\raisebox{-.4ex}{\scriptsize HF}( $E$\raisebox{-.4ex}{\scriptsize F}) was the only one fitting parameter of the DOMP. It was chosen in such a way that the potential used would lead to the experimental value of nucleon binding energy for the bound state under consideration in A+1-nucleon  system.

\section{Calculation of ANCs}

	For brevity, in the following text we will use only one shell-model quantum number $j$ to denote the state, omitting the quantum numbers $n$ and $l$. For the systems we have considered, $j$ coincides with the total angular momentum of the nucleus.

Within the optical model, it is assumed that the overlap function $I_{j}(r)$ and the single-particle function $u_{j}$, which is a solution to the Schr\"odinger equation, have the same radial dependence and differ only in normalization. Hence, according to Eqs. \eqref{baru} and 
\eqref{spectr}
\begin{eqnarray} \label{spectr2}
I_{j}(r)=S_{j}^{1/2}\bar u_{j}/r. 
\end{eqnarray}
At $r\to\infty$ $\bar u_{j}(r)$ behaves as follows
\begin{eqnarray} \label{asympt2}
\bar u_{j}(r) \longrightarrow b_{j}\sqrt{\frac{2\kappa r}{\pi}}K_{l+1/2}(\kappa r)\quad \mbox{for neutrons},\\ \nonumber
\bar u_{j}(r) \longrightarrow b_{j}W_{-\eta,l+1/2}(2\kappa r) \quad \mbox{for protons},
\end{eqnarray}
where $b_{j}$ is called the single-particle ANC.
Comparing Eqs. \eqref{asympt2}  with \eqref{asympt} and \eqref{asympt1} and using \eqref{spectr2} we get a connection between the true ANC $C_{j}$ and $b_{j}$:
\begin{eqnarray} \label{spectr3}
C_{j}=S_{j}^{1/2}b_{j}.
\end{eqnarray}
In this paper, the values of $b_{j}$ are found from Eqs. \eqref{asympt2} from the asymptotic values of the function 
$\bar u_{j}(r)$ obtained by solving the Schr\"odinger equation with the DOMP. After this, the ANC $C_{j}$ is determined from \eqref{spectr3}  using the spectroscopic factor $S_j$ calculated according to \eqref{spectr1} with the same DOMP. 

Let's introduce the notation $b_j(r)$:
\begin{eqnarray} \label{b}
b_j(r)=\frac{\bar u_j(r)}{\sqrt{2\kappa r/\pi}K_{l+1/2}(\kappa r)}\quad \mbox{for neutrons}, \\ \nonumber
b_j(r)=\frac{\bar u_j(r)}{W_{-\eta,l+1/2}(2\kappa r)} \quad \mbox{for protons}.
\end{eqnarray}
From the calculations carried out in this paper it follows that for all the nuclear states considered the value of $b_j(r)$ is constant with high accuracy at $r>8$ fm:
\begin{eqnarray} \label{b1}
b_j(r)|_{r>8\,\mathrm{fm}}=\mbox{const} =b_j.
\end{eqnarray}
As an example, Fig. 1 shows the radial dependence $b_{7/2}(r)$ for the ground state of $^{41}$Ca. For all other systems considered, the corresponding figures look similar to Fig. 1, and we do not present them.

\begin{figure}[h!]
\centering
\includegraphics[scale=1]{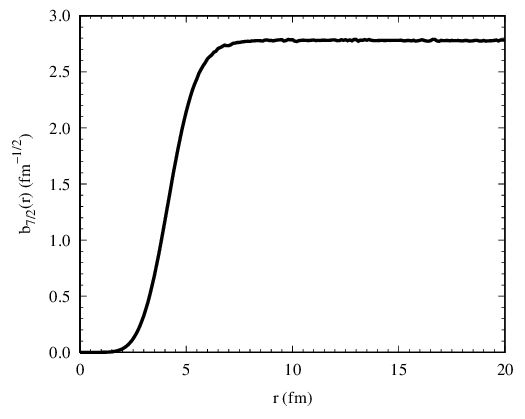} 
\caption {Function $b_{7/2}(r)$ for $^{41}$Ca}
\label{fig1}
\end{figure}

The ANC calculations for the separation of the outer nucleon were carried out within the framework of the approach described above for two pairs of mirror nuclei: $^{17}$O-$^{17}$F and $^{41}$Ca-$^{41}$Sc. Calculations were performed for both the ground and first excited states of nuclei (except for $^{41}$Sc nucleus which has no excited bound  states). All the processes considered are of the type $(A+1)\to A+n(p)$, where the nucleus $A$ is in the ground state. To determine $b_j$, Eq. \eqref{b1} at $r=15$ fm was used. To determine the ANC, it is necessary to describe the nucleon separation energy as accurately as possible. That's why we found two different values of    $V_{HF}(E_F)$, which provided an agreement within 0.5\% of the calculated single-particle energy $E_{nlj}$ of the corresponding nucleon bound states with the experimental separation energy (with the opposite sign) for the ground and excited states in the nuclei under consideration. Looking ahead, we note that resulting two $V_{HF}(E_F)$ values differ one from another by 1-3\%.

\subsection{Calculation of ANCs for $^{41}$Ca and $^{41}$Sc}
	Let's start with the $^{41}$Ca nucleus. For this nucleus, by adjusting the parameter $V_{\textrm{HF}} (E_{\textrm{F}})$ of the HF component, the experimental binding energies of neutron $\varepsilon_{7/2}=8.36$ MeV and $\varepsilon_{3/2}=6.41$ MeV for the ground ($^{41}$Ca$(7/2^-:0\;\mathrm{MeV})$) and the first excited ($^{41}$Ca$(3/2^-;1.95 \:\mathrm{MeV})$) states respectively were reproduced accurate enough. For these values, the solution of the Schr\"odinger equation gives $b_{7/2}=2.77$ 
fm$^{-1/2}$ (see Fig. 1), and Eq. \eqref{spectr1} leads to $S_{7/2}=0.806$. As a result, we obtain $C_{7/2}(\mathrm{Ca})=\sqrt{S_{7/2}}\,b_{7/2} =2.49$ fm$^{-1/2}$. For the $3/2^-$ state $b_{3/2}=8.77$ fm$^{-1/2}$, $S_{3/2}=0.835$, and $C_{3/2}(\mathrm{Ca})=8.02$ fm$^{-1/2}$.

$^{41}$Sc nucleus has only one bound state $7/2^-$ with the binding energy of the outer proton $\varepsilon_{7/2}=1.085$ MeV. The DOMP
 adjusted to describe this energy results in  $b_{7/2}=20.82$ fm$^{-1/2}$, $S_{7/2}=1.049$, and  $C_{7/2}(\mathrm{Sc}) =21.35$ fm$^{-1/2}$.

\subsection{Calculation of ANCs for $^{17}$O and $^{17}$F}
 For mirror nuclei $^{17}$O and $^{17}$F, the ANCs were calculated for the ground ($5/2^+$) and first excited ($1/2^+$) states. The experimental  energies of outer nucleon separation from these states are $\varepsilon_{5/2}=4.14$ MeV, $\varepsilon_{1/2}=3.27$ MeV and $\varepsilon_{5/2}=0.600$ MeV, $\varepsilon_{1/2}=0.105$ MeV for $^{17}$O and $^{17}$F, respectively. Solving the Schr\"odinger equation, these separation energies were reproduced by adjusting the parameter $V_{\textrm{HF}}(E_\textrm F)$ analogously to the case of $^{41}$Ca and $^{41}$Sc. Using Eq. \eqref{spectr1} leads to the following results: for $^{17}$O $b_{5/2}=0.929$ fm$^{-1/2}$, $S_{5/2}=0.842$, and $C_{5/2}(\mathrm{O})=0.853$ fm$^{-1/2}$; $b_{1/2}=3.19$ fm$^{-1/2}$, $S_{1/2}=0.892$, and $C_{1/2}(\mathrm{O})=3.01$ fm$^{-1/2}$. For $^{17}$F $b_{5/2}=0.984$ fm$^{-1/2}$, $S_{5/2}=0.858$, and $C_{5/2}(\mathrm{F})=0.912$ fm$^{-1/2}$; $b_{1/2}=83.9$ fm$^{-1/2}$, $S_{1/2}=0.921$, and $C_{1/2}(\mathrm{F})=80.5$ fm$^{-1/2}$. It should be noted that knowledge of the ANCs for $^{17}$F is important for determining the cross section of $^{16}$O$(p,\gamma)^{17}$F reaction, which is part of the CNO cycle of hydrogen burning in stars.

\section{Summary and discussion}
	Our results presented in the previous section are shown in Table \Ref{tabl1}. Let us compare them with the results of other works.

\begin{table}[htbp]
\caption{ANCs for $^{17}$O, $^{17}$F, $^{41}$Ca, $^{41}$Sc} 
\label{tabl1}
\begin{center}
\begin{tabular}{l c c c }
\hline
Nucleus & $b_j$, fm$^{-1/2}$ & $S_j$ & $C_j$, fm$^{-1/2}$  \\ [1ex]
\hline 
$^{17}$O$(5/2^+)$ & 0.929 & 0.842 & 0.853  \\ 

$^{17}$O$(1/2^+)$ & 3.19 & 0.892 & 3.01   \\ 

$^{17}$F$(5/2^+)$ & 0.984 & 0.858 & 0.912   \\ 

$^{17}$F$(1/2^+)$ & 83.9 & 0.921  & 80.5 \\

$^{41}$Ca$(7/2^-)$ & 2.77 & 0.806 & 2.49 \\

$^{41}$Ca$(3/2^-)$ & 8.77 & 0.835 & 8.02 \\

$^{41}$Sc$(7/2^-)$ & 20.82 & 1.049 & 21.35 \\
\hline 
\end{tabular}
  
\end{center}
\label{table1}
\end{table}
In Ref. \cite{Pang}, the value $C_{7/2}(\mathrm{Ca})=2.89$ fm$^{-1/2}$ was obtained by analyzing reaction $^{40}$Ca$(d,p)^{41}$Ca within the DWBA.  Analysis of the same $(d,p)$ reaction using various optical models, including DOM, yields $C_{7/2}(\mathrm{Ca})$ in the range 1.67--2.24 fm$^{-1/2}$ \cite{Nguyen}. 

The ANCs for removing the nucleon from $^{17}$O and $^{17}$F were determined in a number of works by various methods. In Ref. \cite{Huang} from the analysis of data on the radiative capture of nucleons on $^{16}$O, it was obtained $C_{5/2}(\mathrm{O})=0.90$ fm$^{-1/2}$, $C_{1/2}(\mathrm{O})=3.01$ fm$^{-1/2}$, $C_{5/2}(\mathrm{F})=0.91$ fm$^{-1/2}$, and $C_{1/2}(\mathrm{F})=77.2$ fm$^{-1/2}$.  Extrapolating the phase-shift analysis data on elastic scattering of nucleons on $^{16}$O gives $C_{1/2}(\mathrm{O})=2.23 \pm 0.30$ fm$^{-1/2}$ \cite{BKMS4} and $C_{5/2}(\mathrm{F})=0.88$ fm$^{-1/2}$,  $C_{1/2}(\mathrm{F})=95.5$ fm$^{-1/2}$ \cite{BKMS3}.  Analytical continuation of the experimental differential cross section of the $^{16}$O(d,p)$^{17}$O reaction leads to $C_{1/2}(\mathrm{O})=3.34 \pm 0.14$ fm$^{-1/2}$ \cite{BS24}. The values of the ANCs for $^{17}$F were also extracted by various experimental data in the works \cite{Art,Art1,Ryberg}. The ANCs obtained in these works are often referred to as experimental. They lie in the intervals $C_{5/2}(\mathrm{F})=0.90-1.1$ fm$^{-1/2}$ and $C_{1/2}(\mathrm{F})=73-81$ fm$^{-1/2}$. 

The ANCs considered in the present paper were treated  in a number of theoretical studies. The calculations within the source-term approach \cite{Timof} gave $C_{5/2}(\mathrm O)=0.67$ fm$^{-1/2}$,  $C_{1/2}(\mathrm{O})=1.33$ fm$^{-1/2}$, $C_{5/2}(\mathrm{F})=0.76$ fm$^{-1/2}$, $C_{1/2}(\mathrm{F})=77.2$ fm$^{-1/2}$,  $C_{7/2}(\mathrm{Ca}) =2.10$ fm$^{-1/2}$, $C_{3/2}(\mathrm{Ca}) =4.14$ fm$^{-1/2}$ and $C_{7/2}(\mathrm{Sc}) =17.0$ fm$^{-1/2}$. Usage of the continuum-shell-model approach \cite{Okolow} resulted in $C_{5/2}(\mathrm O)=0.81$ fm$^{-1/2}$,  $C_{1/2}(\mathrm{O})=2.79$ fm$^{-1/2}$, $C_{5/2}(\mathrm{F})=0.88$ fm$^{-1/2}$, and $C_{1/2}(\mathrm{F})=73.7$ fm$^{-1/2}$. 

It follows from the above that our results are consistenet with the experimental (or emperical) values of considered ANCs, which are characterized by a noticeable spread. Particularly good agreement is achieved with the results of Ref. \cite{Huang}, which is specifically aimed at extracting ANCs from an experiment. There is also good agreement with the theoretical results of Ref. \cite{Okolow}. Note that our ANCs systematically exceed the theoretical values of Ref. \cite{Timof}. However, the ratios of our ANC values for mirror states are close to the corresponding ratios from \cite{Timof}. These ratios are also consistent with the general formulas for such ratios presented in Ref. \cite{MukhBlokh}.   

For accurate calculation of  ANCs in our approach, it is important that the potential used reproduces the experimental binding energies  precise enough. Lets consider  the ANC for the ground $7/2^-$ state of $^{41}$Ca ($\varepsilon_{7/2} = 8.36$ MeV) for the example. If  the potential reproduces binding energy of not  the ground state but the excited $3/2^{-}$ state, then it leads to $\varepsilon_{7/2} = 
9.88$ MeV and $C_{7/2}(\mathrm{Ca}) = 3.53$ fm$^{-1/2}$ instead of  $C_{7/2}(\mathrm{Ca}) = 2.49$ fm$^{-1/2}$. That is, a relatively small change in the binding energy leads to a noticeable change in the ANC. The accuracy of the binding energy is especially important for the separation of a weakly bound proton. If, when calculating the ANC for the excited 
$1/2^{+}$ state of $^{17}$F ($\varepsilon_{1/2}=0.105$ MeV) , the potential is adjusted  to reproduce the binding energy of the ground $5/2^+$ state , then such a potential will lead to $\varepsilon_{1/2}=0.023$ MeV, and instead of $C_{1/2}(\mathrm{F})=80.5$ fm$^{-1/2}$ we will get $C_{1/2}(\mathrm{F})=2.17\cdot 10^5$ fm$^{-1/2}$. Such a strong dependence of the result on the binding energy is due to the fact that at $\varepsilon_j\to 0$ the ANC $C_j$ sharply tends to $\infty$ \cite{Okolow}.

It can be concluded that the approach used in this study allows us to successfully determine ANCs for removing nucleons from nuclei. The advantage of this approach is its relative simplicity. The authors plan to apply this approach to other nuclei.

The study was conducted under the state assignment of Lomonosov Moscow State University.

\end{document}